\documentclass[final,3p,times,twocolumn]{elsarticle}
\usepackage{amssymb}

\begin{document}

\begin{frontmatter}

\title{Disorder Induced Effects on the Critical Current Density of Iron Pnictide 
BaFe$_{1.8}$Co$_{0.2}$As$_2$ single crystals}

\author[l1]{M. Zehetmayer}
 \ead{zehetm@ati.ac.at}
\author[l1]{M. Eisterer}
\author[l1]{H. W. Weber}
\author[l2]{J. Jiang}
\author[l2]{J. D. Weiss}
\author[l2]{A. Yamamoto}
\author[l2]{A. A. Polyanskii}
\author[l2]{E. E. Hellstrom}
\author[l2]{D. C. Larbalestier}

\address[l1]{Vienna University of Technology, Atominstitut, 1020 Vienna, Austria}
\address[l2]{National High Magnetic Field Laboratory, Florida State University, Tallahassee, FL
32310, USA}

\begin{abstract}
Investigating the role of disorder in superconductors is an essential part of characterizing the
fundamental superconducting properties as well as assessing potential applications of the material.
In most cases, the information available on the defect matrix is poor, making such studies
difficult, but the situation can be improved by introducing defects in a controlled way, as provided
by neutron irradiation.
In this work, we analyze the effects of neutron irradiation on a Ba(Fe$_{1-x}$Co$_x$)$_2$As$_2$
single crystal. We mainly concentrate on the magnetic properties which were determined by
magnetometry. Introducing disorder by neutron irradiation leads to significant
effects on both the reversible and the irreversible magnetic properties, such as the transition
temperature, the upper critical field, the anisotropy, and the critical current density. The results
are discussed in detail by comparing them with the properties in the unirradiated state.
\end{abstract}


\end{frontmatter}

The recent discovery of a new class of superconducting materials - the so-called iron pnictides -
has initiated a lot of research aiming at understanding the theoretical background and verifying the
potential for applications of these materials. Significant information may be acquired from studying
the effect of disorder, which influences both the reversible and the irreversible superconducting
properties. In this paper, we report on the effects of introducing defects into
BaFe$_{1.8}$Co$_{0.2}$As$_2$ \cite{Sef08,Tan09,Yam09,Eis09} by neutron irradiation. We mainly
concentrate on the irreversible magnetic properties in fields parallel and perpendicular to the
Fe-pnictide layers.

The magnetic properties of a BaFe$_{1.8}$Co$_{0.2}$As$_2$ single crystal with a size of $a \times b
\times c \simeq 1.40 \times 0.70 \times 0.10$\,mm$^3$ were investigated by SQUID and
VSM magnetometry. The sample was investigated in the as-grown state and again after exposing it
to neutrons of various energies in our research reactor to a fluence ($E > 0.1$\,MeV) of $4 \times
10^{21}$\,m$^{-2}$ . Neutrons typically create defects of different sizes by collisions or nuclear
reactions with the lattice atoms. These defects increase the scattering rate and thus will
change the superconducting properties and may serve as new pinning centers.

We started by measuring the transition temperature in a SQUID using the ac-mode with a field
amplitude of 0.1\,mT parallel to $ab$ (parallel to the Fe-As layers). $T_\mathrm{c} \simeq 23.4$\,K
was found in the unirradiated state. Although the slope of (the in-phase) magnetic moment - $m(T)$ -
is quite steep just below $T_\mathrm{c}$, we still observe a small decay (i.e. a small rise of
$|m(T)|$) at 5\,K, indicating minor inhomogeneities, maybe from Co doping. Measuring the Meissner
slope (i.e. $\partial m / \partial H_\mathrm{a}$ in the Meissner state) at 5\,K allows to
approximately determine the sample volume, since geometry effects are insignificant for the chosen
arrangement. Accordingly, the whole sample volume is superconducting.

Upon neutron irradiation, $T_\mathrm{c}$ decreases slightly to about 23.1\,K. The slope at
$T_\mathrm{c}$ becomes somewhat steeper which shows that the radiation induced disorder is
homogeneously distributed. The Meissner slope at 5\,K is slightly flatter now (by $\sim$ 6\%) which
could indicate surface degradation or minor experimental errors (e.g. misalignment). A small decay
of $T_\mathrm{c}$ is quite common in
anisotropic superconductors and can be related to two effects in our sample, i.e. (i) more
disorder provides additional scattering centers for electrons, and (ii) a decay of anisotropy
with irradiation. Further reversible properties were discussed in Ref.~\cite{Eis09}. Briefly, we
found a slight drop of the anisotropy from about 2.9 to 2.5 at high temperatures and
$B_\mathrm{c2}^\mathrm{c}(T)$ that may be extrapolated to roughly 35 - 40\,T at 0\,K.

\begin{figure}
    \centering
	\includegraphics[clip, width = 8.5 cm]{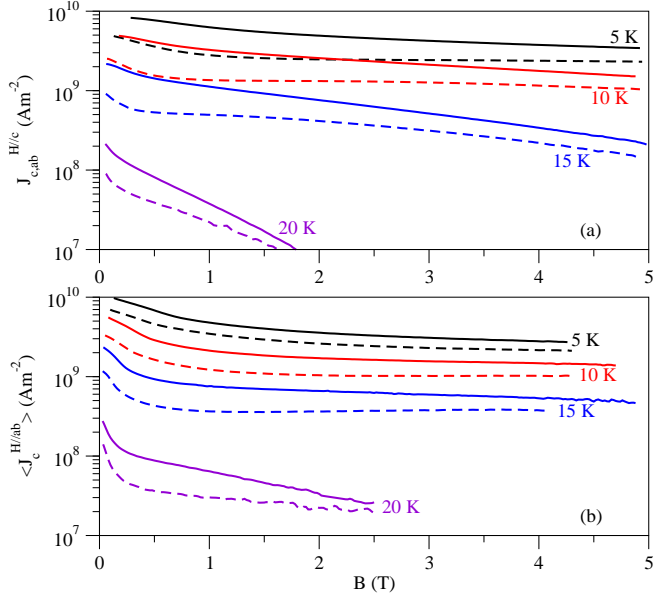}
    \caption{\label{fig:jc} Critical current density for $H_\mathrm{a} \parallel c$
(panel a) and $H_\mathrm{a} \parallel ab$ (panel b) before (dashed lines) and after (solid lines)
neutron irradiation as a function of magnetic induction.}
\end{figure}

The irreversible properties were determined from magnetization loops - $m(H_\mathrm{a})$ -
with $H_\mathrm{a} \parallel ab$ or $H_\mathrm{a} \parallel c$ measured in a VSM  at fields of up to
5\,T and a field sweep rate of 10$^{-2}$ T/s. Since the current flow in the $ab$ plane is assumed
to be isotropic for  $H_\mathrm{a} \parallel c$, $J_\mathrm{c,ab}^{H
\parallel c}$ - the critical current density - can be evaluated by applying Bean's model for
rectangular samples: $J_\mathrm{c,ab}^{H \parallel c}(B) = [|m_\mathrm{i}^{H
\parallel c}(B)|/\Omega][12a/b(3a-b)]$, $\Omega = a b c$,
$m_\mathrm{i}^{H \parallel c}(B)$ denotes the irreversible magnetic moment, i.e. half of the
hysteresis width at fixed $B$ (see Ref.~\cite{Zeh04} for more details).

For $H_\mathrm{a} \parallel ab$ we expect different current densities along $c$ ($J_\mathrm{c,c}^{H
\parallel ab}$)  and $ab$ ($J_\mathrm{c,ab}^{H \parallel ab}$). In this case,  $J_\mathrm{c}$ can
only be calculated from $m$ if one of the components dominates the magnetic properties, namely via
$J_\mathrm{c}^{H \parallel ab}(B) = [|m_\mathrm{i}^{H \parallel ab}(B)|/\Omega][4/s]$, where $s$ is
the sample dimension perpendicular to $H_\mathrm{a}$ and $J_\mathrm{c}^{H \parallel ab}$. Since we
cannot predict the $J_\mathrm{c}$ relations, we apply the isotropic method also in this case
($H_\mathrm{a}\parallel ab$) resulting in an averaged value $\langle J_\mathrm{c}^{H \parallel ab}
\rangle$ which should be somewhere between $J_\mathrm{c,c}^{H \parallel ab}$ and $J_\mathrm{c,ab}^{H
\parallel ab}$.

\begin{figure}
    \centering
	\includegraphics[clip, width = 8.5cm]{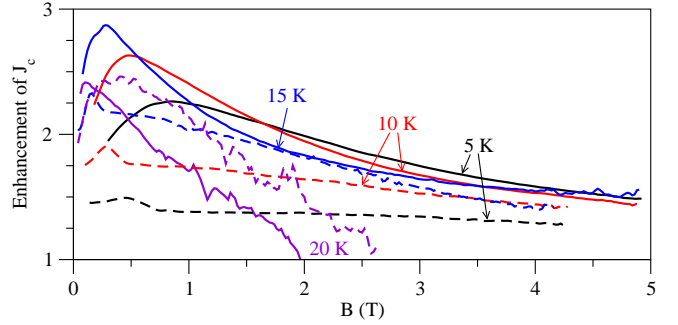}
    \caption{\label{fig:Enhancement} Ratio of critical currents after
 and before neutron irradiation for $H \parallel c$ (solid lines) and $H \parallel ab$
(dashed lines) as a function of applied field at different temperatures.}
\end{figure}

Figure~\ref{fig:jc}a shows results on $J_\mathrm{c,ab}^{H \parallel c}(B)$ at temperatures from 5 to
20\,K. The dashed line indicates the as-grown and the solid line the irradiated state. The radiation
induced enhancement of $J_\mathrm{c}(B)$, i.e. $J_\mathrm{c}^\mathrm{irrad.} /
J_\mathrm{c}^\mathrm{as-grown}$,  is illustrated in Fig.~\ref{fig:Enhancement} by the solid lines. A
maximum enhancement of roughly 3 is found at 15\,K and $\sim0.5$\,T. The effect clearly
decreases with increasing field and with decreasing temperature. A pronounced maximum at small
fields is found at all temperatures, which presumably is the signature of a more ordered flux line
phase in the as-grown state at these fields. The newly created pinning centers are held responsible
for the enhancement, since only minor changes in the reversible properties were found (e.g.
$T_\mathrm{c}$, coherence length, and anisotropy \cite{Eis09}). Thus BaFe$_{1.8}$Co$_{0.2}$As$_2$ is
another material (such as many cuprates and MgB$_2$), in which neutron irradiation creates
defects, that are well suited for flux pinning.

Results on $\langle J_\mathrm{c}^{H \parallel ab} \rangle$ are presented in Fig.~\ref{fig:jc}b. At
low fields, the values are slightly higher than that of panel a (up to $\sim40$\,\%) and the
irreversibility field at high temperature is obviously enhanced (which may be mainly a consequence
of the higher upper critical field in this direction). As already mentioned, the interpretation of
$\langle J_\mathrm{c}^{H \parallel ab} \rangle$ is not straightforward. Note that the front of
$J_\mathrm{c,ab}^{H \parallel ab}$ penetrates into the sample in $c$ direction, and that of
$J_\mathrm{c,c}^{H \parallel ab}$ along $ab$, and $c \simeq 0.1 (ab)^{1/2}$ ($(ab)^{1/2}$ indicates
the mean sample size parallel to the $ab$ plane, i.e. the Fe-As layers). Thus as long as the
currents in $ab$ direction are not much higher than those in $c$ direction (roughly for
$J_\mathrm{c,ab}^{H \parallel ab} < 5 J_\mathrm{c,c}^{H \parallel ab}$) the $ab$ currents dominate
the magnetic properties and we can calculate that component giving: $J_\mathrm{c,ab}^{H \parallel
ab} \simeq \langle J_\mathrm{c}^{H \parallel ab} \rangle$. Similarly, the $c$ component dominates
for roughly $J_\mathrm{c,c}^{H \parallel ab} < 20 J_\mathrm{c,ab}^{H \parallel ab}$, which leads to
$J_\mathrm{c,c}^{H \parallel ab} \simeq 0.1 \langle J_\mathrm{c}^{H \parallel ab} \rangle$ (the
factor 0.1 follows again from the ratio of the sample dimensions). Thus, we may deduce that
$J_\mathrm{c,c}^{H \parallel ab}$ cannot be lower than about $0.1 \langle J_\mathrm{c}^{H
\parallel ab} \rangle$ (e.g. $7 \times 10^8$Am$^{-2}$  at 5\,K and low fields) which is
approximately 0.1$J_\mathrm{c,ab}^{H \parallel c}$. The minimum case would also imply strongly
anisotropic pinning for $H_\mathrm{a} \parallel ab$, similar to the highly
anisotropic cuprates due to a large modulation of the superconducting order parameter perpendicular
to the planes. Such modulations are not excluded in our sample, since the pnictides have also a
layered structure.

It seems more plausible, however, that the current anisotropy is not very large since the anisotropy
of the coherence length is only 2 - 3 (see also \cite{Tan09}). Also recent STM studies \cite{Yin09}
on flux line pinning led to the conclusion that pancake vortices do not exist in this material. Thus
$\langle J_\mathrm{c}^{H \parallel ab} \rangle$ might rather represent $J_\mathrm{c,ab}^{H \parallel
ab}$. In this case $J_\mathrm{c,ab}^{H \parallel ab}$ would be slightly higher than
$J_\mathrm{c,ab}^{H \parallel c}$ but the differences almost disappear upon neutron irradiation,
since $\langle J_\mathrm{c}^{H \parallel ab} \rangle$ increases only by a factor of about 1.5 - 2.2
at 5 - 15\,K. In contrast to $H_\mathrm{a} \parallel c$, the enhancement is almost constant with
field within this temperature range as shown in Fig.~~\ref{fig:Enhancement}. A more pronounced field
dependence is only found at 20\,K.

In summary, we found $J_\mathrm{c,ab}^{H \parallel c}$ values of up to about $5 \times
10^9$Am$^{-2}$ (5\,K) in a BaFe$_{1.8}$Co$_{0.2}$As$_2$ single crystal, which increases by up to a
factor of 3 upon neutron irradiation. For $H \parallel ab$ we showed that  $J_\mathrm{c,c}^{H
\parallel ab}$ cannot be lower than about 0.1 $J_\mathrm{c,ab}^{H \parallel c}$, but much smaller
differences are more likely. The enhancement of the irreversible magnetic properties upon
neutron irradiation is lower for this field direction.

This work was supported by the Austrian Science Fund under contract 21194.

\end{document}